# Deploying SIP-based Mobile Exam Application onto Next Generation Netwok testbed

Ahmed Barnawi, Nadine Akkari, Muhammad Emran, Asif Irshad Khan
Faculty of Computing and IT
King Abdulaziz University
Jeddah, Saudi Arabia
ambarnawi@kau.edu.sa

*Abstract*— over the past few years, mobile operators are faced with enormous challenges. Of such challenges, evolved user demands on personalized applications. Telecommunications industry as well as research community have paid enormous attention to Next Generation Networks (NGN) to address this challenge. NGN is perceived as a sophisticated platform where both application developers and mobile operators cooperate to develop user applications with enhanced quality of experience. The objective of this paper is two folds, first we present an introduction to state-of-the-art NGN testbed to be developed at KAU, and second we provide initial analysis for deploying a mobile application on top of the testbed.

## I. INTRODUCTION

In light of the most recent development in mobile telecommunication market, the concept of mobile application have evolved to become one of most important features in the futuristic mobile communications systems. Recent mobile personalized applications must meet anywhere/anytime/reliable demands set by users. According to standardization body, third Generation partnership Project (3GPP), Virtual Home Environment (VHE), the legacy concept of mobile applications, is defined as a system concept for personalized service portability across network boundaries and between terminals.

Next Generation Networks (NGN) plays central rule to bring together the users, operators, and application developers. From an end-user's point of view, NGN should consistently present him/her with the same personalized services whatever the network that serves him/her and whatever the terminal technology he/she uses. From applications developers' point of view, NGN offers resources and interfaces "APIs" through which the applications are deployed. And from operators' point of view, NGN is concerned with provisioning and guaranteeing end to end (e2e) Quality of Service (QoS) in the context of all-IP heterogeneous network.

The complexity of today's data communications networks necessitates complete, realistic and sophisticated testing playground for verifying and validating their functionalities. Testing in production and commercial networks is typically forbidden since they present a high degree of risk factors for service availability. Therefore, the optimum objective for our research group is the implementation of NGN testbed to be used as a platform for next generation mobile application development.

IP Multimedia Subsystem (IMS) is considered as the cornerstone for NGN. IMS is best described as the glue between the "global" applications world (Internet) and the mobile world. The IMS was designed to make it easy for third party developers to deploy their applications over mobile networks. According to the standards, IMS is defined in the form of reference architecture to enable delivery of next-generation communication services of voice, data, video, wireless, and mobility over an Internet Protocol (IP) network.

Signaling in IMS network is based on a Session Initiation Protocol (SIP). The SIP based architecture provides a multiservice environment with multimedia capabilities. IMS contains Home Subscriber Server (HSS), which is the central storage area for user-related information such as his/her security related information or the service to which the user is subscribed to. It is also consists of the Serving Call Session Control Function (S-CSCF) which acts as the central node of the signaling plane. S-CSCF on one hand is connected to the Application Server that hosts the application and on the other it is connected to HSS and the mobile IMS either through the Proxy CSCF (P-SCSF) if the client resides in its own area of serving or Interrogator CSCF (I-SCSF) if a client is being served by another S-CSCF.

Although IMS was designed to make it easy to develop applications, good knowledge of the IMS network architecture and the underlying Internet protocols is still needed to develop IMS applications [14]. In this paper we highlight our first research findings. The objectives of our research project are to first set up an NGN testbed and second to develop a mobile application on top of the testbed. The application we have chosen to develop is automatic mass production examination system. This system enables students to receive their test on their portable mobile devices through their current access network. The IMS will be used as the control network for our proposed system application which requires user authentication in addition to special security measures related to exam provisioning, exam solving and grading phases. Thus the IMS components responsible to provide these functionalities will be described in details together with the Session Initiation

Protocol SIP used to exchange control messages among the designated IMS components.

The reminder of this paper is organized as follows. In section II, we highlight the IMS components and functionalities. In section III, we give a brief of SIP signaling.

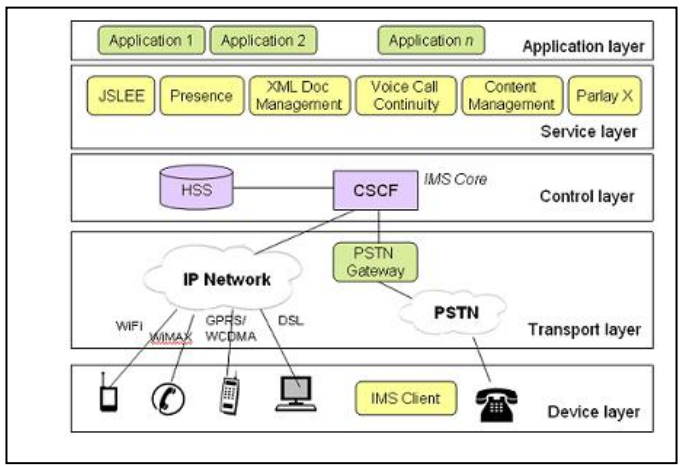

Figure 1. IMS architectural diagram, Source [2]

In section IV, we explain our IMS based testbed architecture. In section V, we introduce the model of the examination system and give an example on SIP- based signaling and analyse the flow of traffic between IMS components. In VI, we conclude and present the future work.

## II. IMS ARCHITECTURE AND COMPONENTS

The 3GPP defines IMS as an architecture framework for delivering multimedia services for both wireless and fixed access technologies based on the Internet Protocol (IP) [1]. SIP has emerged as the vital technology for controlling communication in IP-based Networks. The IMS platform is based on layered architecture design that facilitates standards-based communication to application services as shown in Figure 1.

The IMS core network consists of mainly Transport layer, Control layer and Application layer. The device layer refers to the IMS client applications and devices, while service layer is mainly used to provide added functionality to IMS core network. The IMS main layers characteristics and functions are as follows:

- Transport layer: This layer is designed to provide network access. Initiation and termination of SIP sessions is done by this layer. IMS devices connect to the IP network in the transport layer using different technique including Cable, WiFi, DSL, modem, GPRS, and WCDMA etc.

- Control layer: This is the middle layer which mainly contains CSCF and HSS. Routing and generating billing details for the use of the network are the main responsibility of this layer.

- Application layer: This layer allows service providers to offer different multimedia services to its users. The application servers are in charge of hosting and executing the services and provide interface against the control layers using the SIP protocol. There may be several application servers providing different multimedia services. Presence server, Instant Messaging Server, and Group List Management Server are some core application servers.

The IMS main Components are:

- CSCF: The Call State Control Function (CSCF) is the heart and soul of the IMS. SIP (Session Initial Protocol) is used as signaling protocol for establishing, controlling, modifying and terminating sessions between two or more of the SIP routing machinery. CSCF can be further divided into 3 subcomponents mainly P-CSCF, I-CSCF, S-CSCF.

- The Proxy –CSCF (P-CSCF): is the first point of contact for a user with the IMS and act as an outbound/inbound SIP proxy server. This means that all the requests initiated by the IMS terminal or destined for the IMS terminal traverse the P-CSCF. The P-CSCF includes several functions, some of which are related to security. Since SIP is a text based protocol and sometimes SIP message can be large so the P-CSCF also includes a compressor and a de-compressor of SIP messages using SigComp, which reduces the round-trip over slow radio links. It may also include a PDF (Policy Decision Function), which authorizes media plane resources e.g. quality of service (QoS) over media plane.

- Interrogating-CSCF (I-CSCF): I-CSCF is used to conceal network details from other operators, determining routing within the trusted domain and thus helps to protect the S-CSCF and the HSS from unauthorized access by other networks.

- Serving-CSCF (S-CSCF). The S-CSCF acts as a registrar. It controls subscriber's service (handling registration processes, making routing decisions and maintaining session states, etc) on every session that the user initiates.

- The Home Subscriber Server (HSS): Is the master data storage for all subscribers and service related data of the IMS. The main data stored include user identities, registration information, location of the subscriber device, the services a subscriber is allowed to access and other service-triggering information.

- Application Server (AS): AS is not a part of IMS Core, AS is a SIP unit that hosts and executes services depending upon the services subscribed to and invoked by the user. The ASs offer APIs like SIP servlet, Parlay for application execution.

- The PSTN gateway: PSTN or any other circuit-switched network interacts with IMS through the PSTN gateway. It provides an interface toward a circuit-switched network, allowing IMS terminals to make and receive calls to and from the PSTN (or any other circuit-switched network).

- The Breakout Gateway Control Function (BGCF): Selects which MGCF a call should go through to reach PSTN. If the call goes to the PSTN of another sub-domain, the BGCF of this sub-domain will be selected.

- Media Gateway Control Function (MGCF): To manage calls from or to legacy platforms, calls are routed through the Media Gateway Control Function (MGCF) and passed to the core IMS network CSCF using SIP protocols.

- MGW (Media Gateway): All media processing functions requiring to process calls to/from the PSTN is done by MGW.

### III. SIP SIGNALING [7]

Session Initiation Protocol (SIP) is designed to provide signaling and session management for voice and multimedia services over packet switched networks. It is a peer-to-peer protocol which provides the functionality of intelligent endpoints and distributed call control, such as H.323. SIP supported Gateways do not do not depend on a call agent, although the protocol does define several functional entities that help SIP endpoints locate each other and establish a session [7]. SIP helps in the convergence of voice and data services to meet the emerging demand of the IP Applications over the cellular devices. With its foundation in Internet protocols, SIP provides the ability to integrate traditional voice services with Web-based data services, including self-based provisioning, instant messaging, presence, and mobility services [8].

SIP is the fundamental protocol which works with the IMS technology. It is a lightweight protocol, and offers a vehicle for rich media interworking between end points. SIP has been used in next generation-networks (NGN) since the late 1990s. In 2000, 3GPP adopted SIP as the protocol standard and as the framework for the IMS control. As part of the IMS model, SIP extensions have been added to accommodate a rich service model that assumes multiple applications interacting with each user session [9].

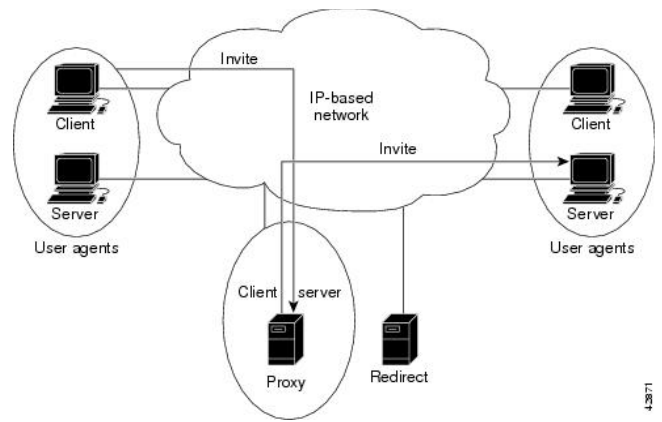

Figure 2.  SIP Request through a Proxy Server, [7]

#### A. SIP flow with Proxy Server

In Fig. 2, a proxy server is used, the caller user agent sends an INVITE request to the proxy server, the proxy server resolve the path, and then forwards the request to the called party [10]. In Fig. 3, called party responds to the proxy server, which in turn, forwards the response to the caller [10]. The proxy server forwards the acknowledgments of both parties. A session is then established between the caller and called party. As shown in Fig. 4, real-time Transfer Protocol (RTP) is used for the communication between the caller and the called party [10].

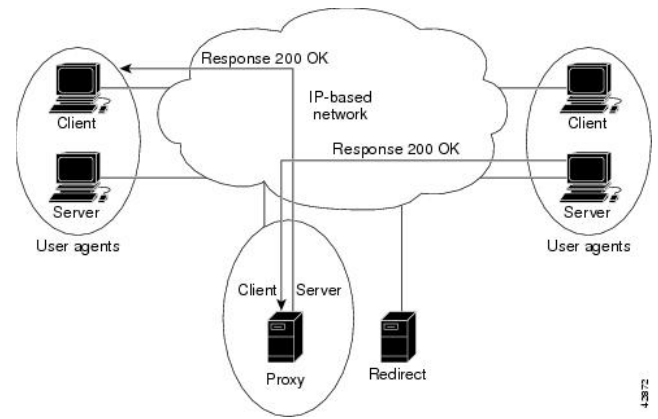

Figure 3.  SIP Responses through a Proxy Server, Source [7]

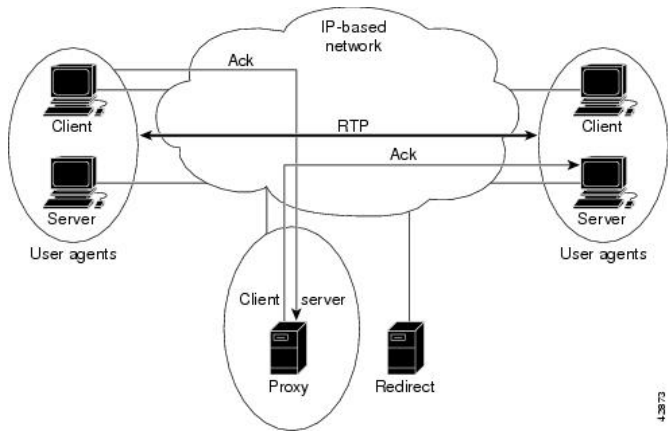

Figure 4.  SIP Sessions through a Proxy Server

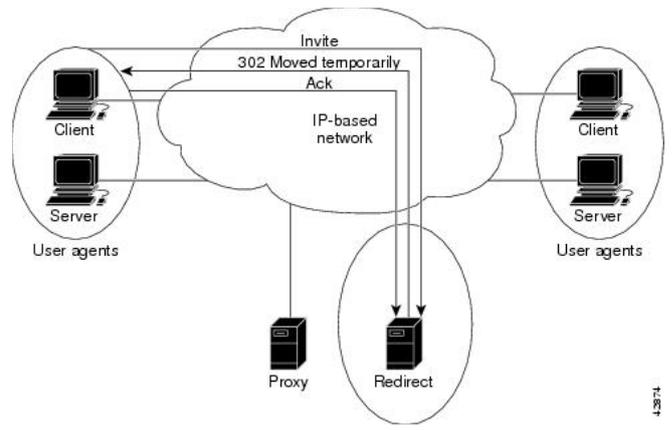

Figure 5.  SIP Request through a Redirect Server

## B. SIP flow with direct Servers

In Fig. 5, a redirect server is used, the caller user agent sends an INVITE request to the redirect server, the redirect server contacts the location server to locate the path to the called party, and then the redirect server sends that information back to the caller. The caller then acknowledges receipt of the information [10].

The caller then sends a request to the device, indicated in the redirection information (which could be the called party or another server that forwards the request to the called party). Once the request reaches the called party, it sends back a response and the caller acknowledges the response. Then finally RTP is used for the communication between the caller and the called party [10], as shown in Fig 6.

## IV. HIGH LEVEL DESIGN OF THE PROPOSED IMS TESBED

The proposed IMS testbed is based on the IMS technology. IMS core will be extended initially with three application servers which would be able to handle the mass examination in KAU. User identification and non-voice communication for submitting, scheduling, conducting and evaluating the student's exams is the core application area to be tested on the proposed IMS testbed.

Fig. 7 shows the services on proposed IMS testbed. These services mapped on the following three servers (1-3) in the proposed high level design of the IMS testbed.

1. This is high end Application Layer Server on which Mass Examination Application will be hosted. Presence Server and optionally IP TV applications can also be installed on this server.

2. This is a high end Core/Control Layer Server; all the core applications will be installed on this server, which includes Open IMS, VOIP and Media Proxy.

3. This is a mid range Transport Layer Server connected with outside world to provide access to the mobile users and will host supporting services for core and application layers. Routing, security and monitoring of the network traffic will also be taken care of by this server.

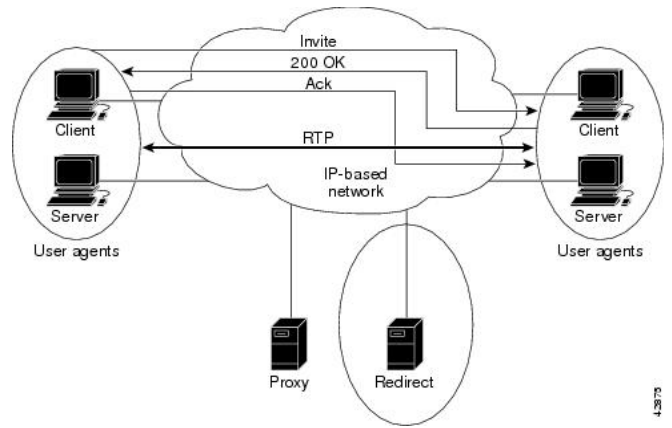

Figure 6.  SIP Sessions through a Redirect Server

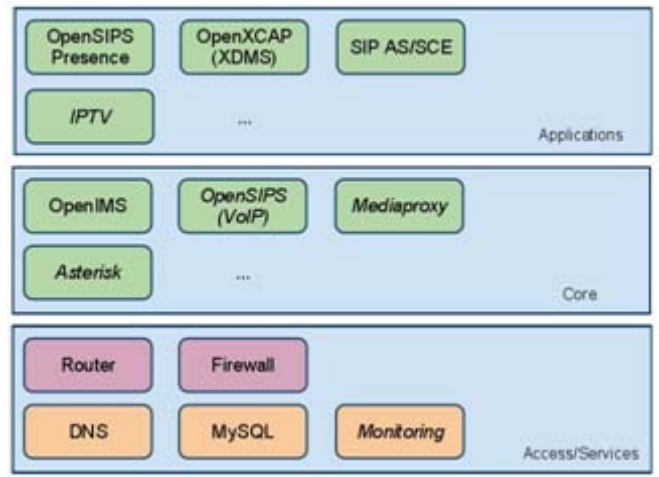

Figure 7.  Services on Proposed High Level Design of IMS Testbed

## V. SIP-BASED PROPOSED EXAM APPLICATION MODEL

The mobile exam scenario consists of an IMS client, web browser, and legacy PSTN/GSM to perform the following operations as illustrated in Fig. 8:

- Exam provisioning: A teacher / professor will enter the exam data through a web page and set up the corresponding schedule through a web- based calendar scheduler. Sending the exam on specified time could be triggered by network based scheduler (steps 1 and 2).

- Determine the group of recipients: Predefined group in XDMS. Accordingly, students are authenticated through pass key based /voice based process (step 3).

- Test submission: Students receive exam announcement via IMS instant message, fill in the exam and submit their exams through their mobile/desktop browser or using voice based or form based (step 3).

- Send results and evaluation: Through voice, SMS, email or instant message to students (steps 4). In addition to sending filled forms back to teachers to be corrected manually or sending generated results after automated evaluation (step 5).

To realize the "mobile exam" use case described in the above scenario, we first set up a minimal solution, which incorporates only the absolutely required components and building blocks, exclusively IP / SIP /NGN based. It could also be extended to include a diversified solution allowing for voice dialogues, integrating legacy phones (mobile / fixed line POTS) or a fully diversified solution, allowing also for speech recognition and SMS. The minimal VHE-NGN testbed requires the following components:

- IMS Clients for mobile and desktop end-devices (Messaging, Presence, VoIP, Video, File sharing, etc.)
- An IMS (3GPP Standard) for session control (e.g. Registration, VoIP, Messaging, Video)
- A Http/SIP Application server (JSR 116/289 Standard)
- An XML Document Management Server (XDMS, OMA Standard)

### A. SIP signaling scenario for user agent subscription

In addition to the above mentioned components, the following IMS components will contribute in the Mobile Exam case as shown in Fig. 9. Accordingly, the AS, HSS, XDMS, CSCF will operate as follows:

- Application server (AS): The exam would be entered by teacher as well as the answers entered by students

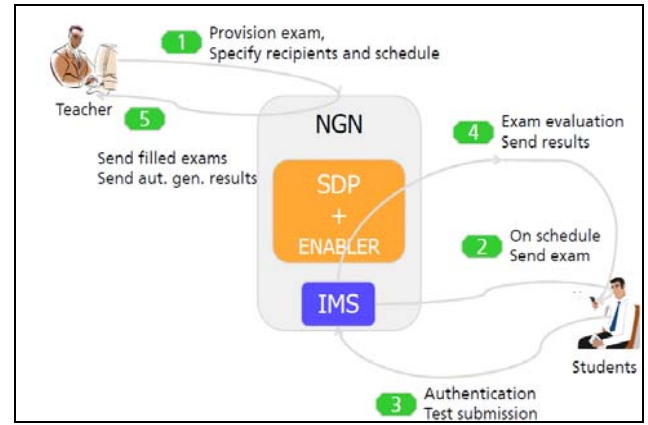

Figure 8. Mobile exam use acse

through Web Page running on Http/SIP AS. In addition, scheduler would be implemented as servlets on AS.

- XML Document Management Server (XDMS): Exam and student group lists would be stored on XDMS. XDMS is a server which keeps all the user information regarding its authentication, applications to which a user can subscribe in the markup language format [11]. The User Agent UA initially sent a Register-message to register in an IMS network (an SCSCF will be assigned). The UA then sends subscribe request to the S-CSCF in order to register to an application or service.

- Home Subscriber Server (HSS): It provides information to the I-CSCF for locating the S-CSCF. It Provides service profile information to the S-CSCF.

- P-CSCF: The P-CSCF is the proxy point for all SIP messages from end-points to the rest of the IMS network. It could be in the home network or may reside in the visited network. The P-CSCF determines what I-CSCF to send SIP messages, which could be an I-CSCF in its own network or another I-CSCF across an administrative domain [12].

- Interrogating-CSCF (I-CSCF): is responsible for finding the S-CSCF at registration. The main function of the I-CSCF is to proxy between the P- and S-CSCF [12].

- Serving-CSCF (S-CSCF): is responsible for interfacing with the Application Servers (AS). When receiving a registration request as a SIP message from an I-CSCF, the S-CSCF will query the HSS via Diameter protocol to register the terminal as being currently served by itself [12]. S-CSCF contacts XDMS to authenticate and subscribe the user to given service. After successful subscription a notification is sent between XDMS to the S-CSCF regarding the successful subscription of the user and finally S-CSCF forward this notification to the user agent UA [12]. The corresponding SIP signaling example among UA, SCSF, and XDMS for mobile user subscription is shown in Fig. 10 as described in [1]. Fig. 11 shows the SIP signaling exchange between CSCF components [13].

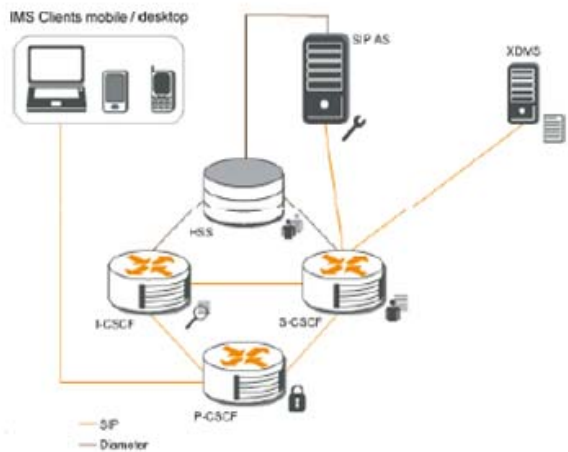

Figure 9. Mobile exam case

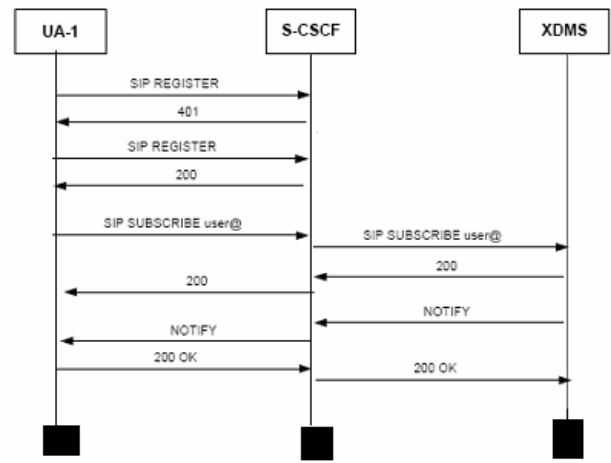

Figure 10. SIP signaling example for mobile user subscription

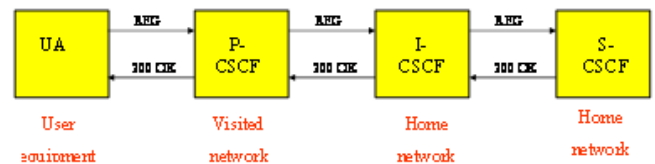

Figure 11. SIP signaling example for CSCF signaling

## VI. CONCLUSION AND FUTURE WORK

In this paper we have present a high level design of future NGN testbed based on IMS technology. The proposed testbed is made of several servers designated for IMS Software components. The testbed will be available for research and business communities to test future mobile applications prior to actual deployment. This will leverage both the developers and operators to develop applications with guaranteed quality of experience. As a case study, we intend to deploy mobile application on top of the testbed. The application chosen is mass examination system. In this paper we have presented some signaling scenarios relating IMS components under this context. We present analysis of SIP-based signaling flow that gives an insight on signaling exchange according to scenarios in the application of concerns.

In future work we will continue with the scope of our project. A lower level design will be designed and system components will be integrated. Parallel to this we will focus on software development life cycle to deploy applications on top of the testbed. Our attention will be paid toward integration of software components and designing testing scenarios. The mass examination application will be the centerpiece application used as proof of concept toward a generic application deployment.